\documentclass[aps,prl,twocolumn, nofootinbib, preprintnumbers, english]{revtex4-1}
\usepackage{amsthm}
\usepackage{amsmath}
\usepackage{amsfonts}
\usepackage{amssymb,epsfig,setspace}
\usepackage{graphicx}
\usepackage{babel}
\usepackage{verbatim}

\begin{document}

\title{Comment on ``Integrability of the Rabi model"}

\author{Alexander Moroz}

\affiliation{Wave-scattering.com}
 
\begin{abstract}
    
\end{abstract}

\maketitle 

\begin{figure}
\begin{center}
\includegraphics[width=\columnwidth]{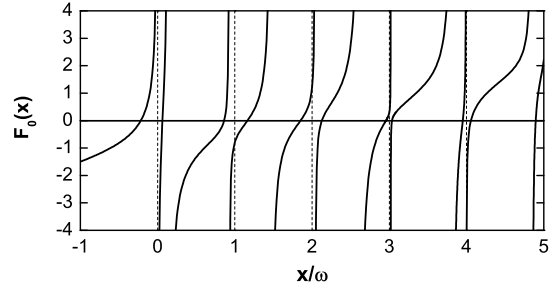}
\end{center}
\caption{\label{fgf0f} Plot of $F_0(x)$ for $g=0.7$, $\Delta=0.4$, and $\omega=1$,
i.e. the same parameters as for $G_\pm(x)$ in Fig. 1 of
Ref. \cite{Br}, shows corresponding 
zeros at $\approx -0.217805$, $6.29563\times 10^{-2}$,
$0.86095$, $1.1636$,  $1.85076$, etc.
}
\end{figure}

In his recent letter \cite{Br}, Braak suggested
that a {\em regular} spectrum of the Rabi model was 
given by the zeros of a transcendental function $G_\pm(x)$
(cf Eqs. (3)-(5) of Ref. \cite{Br}) and  
highlighted the role of the discrete $\mathbb{Z}_2$-symmetry, 
or parity, in determining $G_\pm(x)$.
We show here to the contrary that one can define a transcendental function 
$F_0(x)$ and obtain the regular spectrum 
of the Rabi model as the zeros of $F_0(x)$ (see Fig. \ref{fgf0f})
without ever making use of the underlying 
$\mathbb{Z}_2$-symmetry of the model.

In the latter case 
the task amounts to finding an entire function 
$\phi(z)=\sum_{n=0}^\infty c_n z^n$ which belongs to the Bargmann space 
of analytical functions ${\cal B}$ \cite{Schw} and which coefficients $c_n$
satisfy a three-term recurrence relation 
(Eq. (4) of Ref. \cite{Br}; Eqs. (A.6), (A.8) of Ref. \cite{Schw})
\begin{equation}
y_{n+1} + a_n y_n + b_n y_{n-1}=0,
\label{3trg}
\end{equation}
with $a_n=-f_n(x)/(n+1)$, where $f_n(x)$ is given by Eq. (5) of \cite{Br}
(cf Eq. (A.8) of Ref. \cite{Schw}), and $b_n=1/(n+1)$.
Since $a_n\rightarrow -\omega/(2g)$ and $b_n \rightarrow 0$ 
in the limit $n\rightarrow\infty$,
there exist, in virtue of the theorem by Perron (Theorem 2.2 in Ref. \cite{Gt}),
two linearly independent solutions of the recurrence.
One of the two solutions guaranteed by Perron's theorem is the so-called
{\em minimal} solution that satisfies $\lim_{n\rightarrow\infty} y_{n+1}/y_n =0$
\cite{Gt}. On taking for $c_n$ the minimal solution, 
$\phi(z)$ would belong to ${\cal B}$ 
{\em irrespective} of the value of $x$ and  
of the parameters $g$, $\Delta$, and $\omega$ of the Rabi model \cite{Br}.
The ratios of subsequent terms $c_{n+1}/c_{n}$ of the minimal solution 
are related to continued fractions 
(cf Theorem 1.1 due to Pincherle in Ref. \cite{Gt})
\begin{equation}
r_{n}= \frac{c_{n+1}}{c_{n}} = \frac{-b_{n+1}}{a_{n+1}-} \frac{b_{n+2}}{a_{n+2}-}
\frac{b_{n+3}}{a_{n+3}-}\cdots
\label{rncf}
\end{equation}
Now follows a point of crucial importance. Mathematical theorems 
on three-term recurrence relations in Ref. \cite{Gt} are derived 
assuming $n\geq 1$ in (\ref{3trg}).
On the other hand, physical problems require the recurrence (\ref{3trg}) 
to be also valid for $n=0$ \cite{Schw}.
Given the requirement of analyticity ($c_{-k}\equiv 0$ for $k>1$), 
the three-term recurrence (\ref{3trg}) reduces for $n=0$ 
to an equation involving mere {\em two-terms} and imposes that 
$r_0=c_1/c_0=-a_0=f_0(x)$. However, $r_0$ has been unambiguously 
fixed by (\ref{rncf}), while taking into account (\ref{3trg}) for
$n\geq 1$ only \cite{Gt}, and in general $r_0\neq f_0(x)$ \cite{Schw}. 
It turns out that Eq. (\ref{3trg}) for $n=0$ plays 
for the three-term recurrence (\ref{3trg}) a role 
analogous to the {\em boundary condition}
imposed on the solutions of the Schr\"{o}dinger equation which
enforces quantization of energy levels.
The expansion coefficients $c_0$ and $c_1$ of 
entire function $\phi\in {\cal B}$ defined by the minimal 
solution of (\ref{3trg}) for $n\geq 1$ solve Eq. (\ref{3trg}) 
for $n=0$ if and only if energy parameter $x$ belongs 
to the {\em regular} spectrum of the Rabi model.
The regular spectrum can be thus obtained
as zeros of the transcendental function $F_0(x)\equiv f_0(x)- r_{0}$
(see Fig. \ref{fgf0f}), 
which was deemed to be impossible in Ref. \cite{Br}.
In arriving at the results, one can employ an intimate relation between 
continued fractions and infinite series due to Euler's theorem
and express $r_{0}$ in (\ref{rncf}) 
as $r_{0}=\sum_{k=1}^\infty \rho_1\rho_2\ldots\rho_k$,
where $\rho_l$'s are determined iteratively in terms of the coefficients 
$a_n$ and $b_n$ of (\ref{3trg}) (see recurrence (4.4)-(4.5) in Ref. \cite{Gt}
forming the basis 
of the ``{\em third}" method of Gautschi \cite{Gt}, which has also been used
in our numerical implementation \cite{AMr}).
The latter renders the calculation of 
$F_0(x)$ not more complex than that of
$G_\pm(x)$ in Ref. \cite{Br}.
Thereby we have elucidated and rehabilitated 
a largely misunderstood and
neglected criterion by Schweber \cite{Schw} and
solved it with an efficient numerical implementation, which could,
hopefully, be useful in solving related quantum models.


\end{document}